\documentclass[prl,twocolumn,showpacs,amsmath,amssymb]
{revtex4}

\usepackage[dvips]{color}
\usepackage{graphicx}

\begin{document}

\title{Universal properties of hard-core bosons confined on 
one-dimensional lattices}

\author{Marcos Rigol}
\affiliation{Institut f\"ur Theoretische Physik III, Universit\"at 
Stuttgart, Pfaffenwaldring 57, D-70550 Stuttgart, Germany.}
\author{Alejandro Muramatsu}
\affiliation{Institut f\"ur Theoretische Physik III, Universit\"at 
Stuttgart, Pfaffenwaldring 57, D-70550 Stuttgart, Germany.}

\begin{abstract}
Based on an exact treatment of hard-core bosons confined on one-dimensional 
lattices, we obtain the large distance behavior of the one-particle density 
matrix, and show how it determines the occupation of the lowest natural 
orbital in the thermodynamic limit. We also study the occupation 
$\lambda_{\eta}$ of the natural orbitals for large-$\eta$ at low densities. 
Both quantities show universal behavior independently of the confining 
potential. Finite-size corrections and the momentum distribution function 
for finite systems are also analyzed.
\end{abstract}

\pacs{03.75.Hh, 05.30.Jp}
\maketitle

Trapped atomic gases at very low temperatures became in the past years
a center of attention in atomic and condensed matter physics.
Particularly interesting is the 
case where the dynamics of the system is restricted to one-dimension 
(1D) due to a strong transversal confinement. 
It has been shown recently \cite{olshanii98} that 
in regimes of large positive scattering length, low densities, 
and low temperatures, a quasi-1D gas of bosons behaves as a gas of 
impenetrable particles, i.e., as hard-core bosons (HCB). 
Ultracold Bose gases in 1D have been realized experimentally 
\cite{schreck01}, so that it is expected that 
soon it will be possible to make the 1D HCB gas a physical reality.

The 1D gas of HCB was first introduced theoretically by Girardeau 
\cite{girardeau60}, who also established its exact mapping to a gas of 
noninteracting spinless fermions. Since then, it remained a subject of
recurring attention, and a number of exact results were obtained
for the momentum distribution function (MDF) $n(k)$
and the one-particle density matrix (OPDM) $\rho (x)$
in the homogeneous \cite{lenard64,vaidya79} and the periodic 
\cite{kitanine02} case. 
It was shown that $n(k) \sim |k|^{-1/2}$ for $k \rightarrow 0$, and 
that such a singularity arises due to the 
asymptotic behavior $\rho (x) \sim |x|^{-1/2}$ for large $x$ 
\cite{vaidya79,kitanine02}. Similar results can be obtained using 
the hydrodynamic approximation (bosonization) \cite{cazalilla04}. 

Recently, the attention has been concentrated on the 
ground state properties of the 1D gas of HCB confined by a harmonic 
potential \cite{girardeau01}, as a model for
the experiments. It was found that the 
occupation of the lowest natural orbital (NO) of the trapped system 
scales as $\sqrt{N_b}$, where $N_b$ is the total number of particles, 
as in the homogeneous case (the NO are 
defined as the eigenstates of the OPDM \cite{penrose56}).
The results in 1D are in contrast with higher dimensional 
systems, where a Bose-Einstein condensate (BEC) was proved to exist
\cite{lieb02}, and complement that proof by showing that in 1D only
a quasicondensate is possible. The introduction of an optical lattice 
opens new possibilities to engineer strongly interacting states in the 
trapped gases \cite{greiner02}. Unfortunately, much less is known in the 
lattice case with confinement.

We present here an exact study of trapped HCB on a lattice. 
By means of the Jordan-Wigner transformation \cite{jordan28}
we calculate exactly the Green's function for large 
systems (up to $10^4$ lattice sites). We find that the OPDM 
$\rho_{ij}\sim x^{-1/2}$ ($x = \mid x_i -x_j \mid$) 
for large $x$, irrespective of the confining potential 
chosen, even when portions of the system 
reach occupation $n_i=1$, such that coherence is lost there.  
The power law above is shown to determine the scaling of the occupation 
of the lowest NO in the thermodynamic limit (TL).
In addition we find a power-law decay of the NO occupations 
($\lambda_\eta$) for large-$\eta$ at low densities, 
and show that its exponent is also universal.

We consider HCB on a lattice with a confining potential with the 
following Hamiltonian
\begin{equation}
\label{HamHCB} H_{HCB} = -t \sum_{i} ( b^\dagger_{i} b^{}_{i+1}
+ \text{H.c.} ) + V_\alpha \sum_{i} x_i^\alpha\ n_{i },
\end{equation}
with the addition of the on-site constraints 
$b^{\dagger 2}_{i}= b^2_{i}=0$, $\left\lbrace  
b^{}_{i},b^{\dagger}_{i}\right\rbrace =1$. 
The creation and annihilation operators for the HCB are given by 
$b^{\dagger}_{i}$ and $b_{i}$ respectively, $n_{i }= 
b^{\dagger}_{i}b^{}_{i}$ is the particle number operator, $t$ is the hopping
parameter and the last term in Eq.\ (\ref{HamHCB}) describes an 
arbitrary confining potential, with power $\alpha$ and strength 
$V_{\alpha}$. 

The Jordan-Wigner transformation,
\begin{equation}
\label{JordanWigner} b^{\dag}_i=f^{\dag}_i
\prod^{i-1}_{\beta=1}e^{-i\pi f^{\dag}_{\beta}f^{}_{\beta}},\ \ 
b_i=\prod^{i-1}_{\beta=1} e^{i\pi f^{\dag}_{\beta}f^{}_{\beta}}
f_i \ ,
\end{equation}
maps the HCB Hamiltonian into the one of noninteracting fermions 
\begin{equation}
H_F =-t \sum_{i} (f^\dagger_{i}
f^{}_{i+1} + \text{H.c.})+ V_\alpha \sum_{i} x_i^\alpha
n^f_{i }, 
\end{equation}
with $f^\dagger_{i}$ and $f^{}_{i}$ being creation and 
annihilation operators for spinless fermions, and 
$n^f_{i}=f^\dagger_{i}f^{}_{i}$. 

The Green's function for the HCB can be expressed using 
Eq.\ (\ref{JordanWigner}) as 
\begin{eqnarray}
\label{green1} G_{ij}&=&\langle \Psi^{G}_{HCB}|
b^{}_{i}b^\dagger_{j}|\Psi^{G}_{HCB}\rangle \nonumber \\
&=&\langle \Psi^{G}_{F}| \prod^{i-1}_{\beta=1}
e^{i\pi f^{\dag}_{\beta}f^{}_{\beta}} f^{}_i f^{\dag}_j
\prod^{j-1}_{\gamma=1} e^{-i\pi f^{\dag}_{\gamma}f^{}_{\gamma}}
|\Psi^{G}_{F}\rangle.
\end{eqnarray}
$|\Psi^{G}_{HCB}\rangle$ is the 
ground state for the HCB and $|\Psi^{G}_{F}\rangle$ 
is the ground state for the noninteracting fermions.
The latter is a Slater determinant, i.e., a product of single particle 
states  $|\Psi^{G}_{F}\rangle=
\prod^{N_f}_{\delta=1}\ \sum^N_{\sigma=1} \ P_{\sigma 
\delta}f^{\dag}_{\sigma}\ |0 \rangle,$
with $N_f$ the number of fermions ($N_f=N_b$), $N$ the number of 
lattice sites and ${\bf P}$ is the matrix of the components of 
$|\Psi^{G}_{F}\rangle$. It is easy to see that the action of 
$\prod^{j-1}_{\gamma=1} e^{-i\pi f^{\dag}_{\gamma}f_{\gamma}}$ on 
the fermionic ground state in Eq.\ (\ref{green1}) generates only a 
change of sign on the elements $P_{\sigma \delta}$ for $\sigma \leq j-1$, 
and the further creation of a particle at site $j$ implies the addition 
of one column to ${\bf P}$ with the element $P_{jN_f+1}=1$ and all 
the others equal to zero [the same applies to the action of $
\prod^{i-1}_{\beta=1} e^{i\pi f^{\dag}_{\beta}f_{\beta}} f_i$ on 
the left of Eq.\ (\ref{green1})]. Then the HCB Green's function can 
be calculated exactly as
\begin{eqnarray}
\label{determ}
G_{ij}
&=&\langle 0 | \prod^{N_f+1}_{\delta=1}\ \sum^N_{\beta=1} \ 
P'^{A}_{\beta \delta}f_{\beta} 
\prod^{N_f+1}_{\sigma=1}\ \sum^N_{\gamma=1} \ P'^{B}_{\gamma 
\sigma}f^{\dag}_{\gamma}\ |0 \rangle \nonumber \\
&=&\det\left[ \left( {\bf P}^{'A}
\right)^{\dag}{\bf P}^{'B}\right],
\end{eqnarray}
where ${\bf P'}^{A}$ and ${\bf P'}^{B}$ are the new matrices
obtained from ${\bf P}$ when the required signs are changed and
the new columns added. A proof of the last step 
in Eq.\ (\ref{determ}) can be found in Ref.\ \cite{alejandro}. 
The evaluation of $G_{ij}$ 
is done numerically and the OPDM is given by 
$\rho_{ij}=\left\langle b^\dagger_{i}b_{j}\right\rangle =G_{ij}
+\delta_{ij}\left(1-2 G_{ii} \right)$. The NO ($\phi^\eta_i$) 
can be determined by the eigenvalue equation 
$\sum^N_{j=1} \rho_{ij}\phi^\eta_j= \lambda_{\eta}\phi^\eta_i$, 
with $\lambda_{\eta}$ being their occupations.

\begin{figure}[h]
\includegraphics[width=0.48\textwidth,height=0.32\textwidth]
{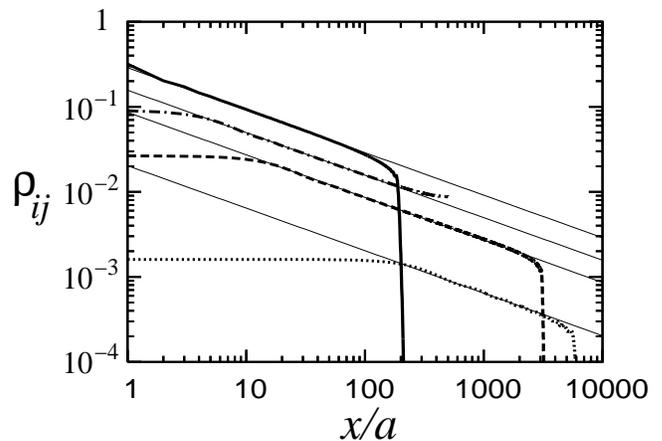}
\caption{OPDM vs $x/a$ ($x = \mid x_i -x_j \mid$) for  
a periodic system with $\rho=9.1 \times 10^{-2}$, 
$N_b$=91 (dashed-dotted line), harmonic traps ($\alpha=2$) with 
$\tilde{\rho}=4.5 \times 10^{-3}$, 
$N_b$=100 (dashed line) and $\tilde{\rho}=2.7$, $N_b$=501 
(thick continuous line, a $n_i=1$ region 
is present), and a trap with power $\alpha=8$, 
$\tilde{\rho}=7.6$ $\times$ $10^{-4}$, $N_b$=11 (dotted line). 
In the trapped cases the abrupt reduction of $\rho_{ij}$ occurs when 
$n_j\rightarrow 0,1$, for $n_i\neq 0,1$ and $i$ chosen arbitrarily. 
Thin continuous lines correspond to power laws $\sqrt{x/a}$.}
\label{Largex}
\end{figure}

We focus next on the large-$x$ behavior of the 
OPDM. For the periodic case ($V_\alpha = 0$) we obtain that for any 
density $\rho\equiv N_b/N\neq 0,1$ the OPDM decays as a power law 
$\rho_{ij} \sim A_{\rho}/\sqrt{x/a}$ for large $x$ (Fig.\ \ref{Largex}), 
where $A_{\rho}$ depends only on the density ($a$ is the 
lattice constant). This behavior was found before by means of 
exact analytical treatments \cite{kitanine02}. In the presence of a 
confining potential, the case relevant for the 
experiments with ultracold atoms, the situation is more complicated 
since the system loses translational invariance and no analytical results 
are available. We first analyze the case where $n_i< 1$ all 
over the system. We find, remarkably, that in this case the OPDM decays 
as a power law $\rho_{ij}\sim A^{\alpha}_{\tilde{\rho}}|x/a|^{-1/2}$ 
for large $x$, i.e., {\em independently} of the local changes 
of the density. (They become relevant only when $n_i,n_j\rightarrow 0$.) 
$A^{\alpha}_{\tilde{\rho}}$ depends on the 
characteristic density of the system $\tilde{\rho}=N_ba/\zeta$ 
and the power $\alpha$ of the confining potential. 
$\zeta=\left( V_\alpha/t\right)^{-1/\alpha}$ is a length 
scale of the trap in the presence of the lattice \cite{rigol03_3}. 
Moreover, the exponent of the OPDM power-law decay does not depend on the 
power $\alpha$ of the confining potential, i.e., it is universal 
(Fig.\ \ref{Largex}). 

\begin{figure}[t]
\includegraphics[width=0.69\textwidth,height=0.457\textwidth]
{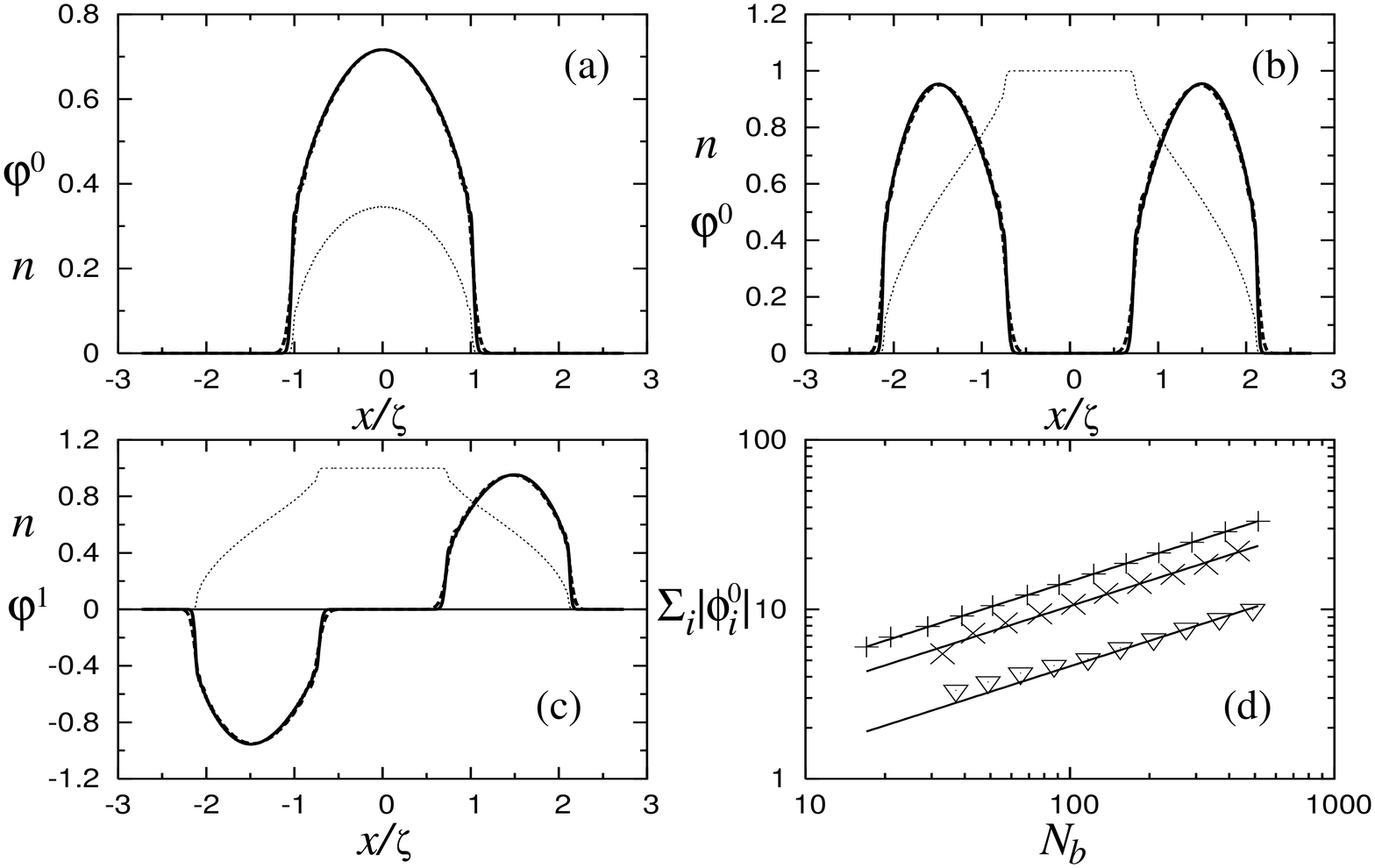}
\caption{Scaled lowest NO and density profiles (thin dotted lines) 
for harmonic traps with (a) $\tilde{\rho}=0.55$, 
$N_b$=101 (continuous line), $N_b$=30 (dashed line); (b),(c) 
$\tilde{\rho}=3.0$, $N_b$=551 (continuous line), $N_b$=167 (dashed line);
(d) shows $\sum_i |\phi^0_i|$ vs $N_b$, in traps with power $\alpha=8$ 
of the confining potential, for $\tilde{\rho}=1.0\ (+),\ 2.0\ (\times),\ 2.25\ 
(\nabla$, a $n_i=1$ region is present); the continuous lines correspond to 
power laws $\sqrt{N_b}$.}
\label{HCB_all}
\end{figure}

A drastic difference between the continuous case and the one with a
lattice is the possibility to build up regions with densities $n_i=1$, 
so that such sites are not any more coupled coherently to the rest. Once 
such regions appear, many NO become occupied with $\lambda_\eta=1$, and all 
the other NO (with $\lambda_\eta\neq 1$) become pairwise degenerated 
since the systems is split in two identical part by the $n_i=1$ plateau 
[Figs.\ \ref{HCB_all}(b),(c)]. Even in this case we find that the OPDM decays 
as a power law $\rho_{ij}\sim A^{\alpha}_{\tilde{\rho}}|x/a|^{-1/2}$ for 
large $x$ in the regions away from $n_i,n_j\neq 0,1$ (Fig.\ \ref{Largex}). 

The universal behavior of the OPDM at large-$x$ above 
shows that although the 1D HCB gas does not exhibit BEC in the TL 
\cite{yang62}, quasi-long-range order is present and a large 
occupation of the lowest NO can occur. In the periodic case the NO 
are plane waves, so that their occupation and the MDF coincide. 
The results for the large-$x$ behavior of the OPDM 
in the periodic case imply that in the TL $n_{k=0}$ scales as
$\sqrt{N_b}$ at constant $\rho$, and $n_k$ diverges as $|k|^{-1/2}$ 
for $k \rightarrow 0$ and $N_b\rightarrow \infty$, 
i.e., in the same way as in the case without a lattice 
\cite{lenard64,vaidya79}.

In the trapped system, due to loss of translational invariance, the NO and 
the MDF do not coincide. To obtain the behavior of the lowest NO in the 
TL, we study how it scales when the strength of the trap (or the number of 
particles) is changed keeping the characteristic density constant. 
In Figs. \ref{HCB_all}(a)-(c) we show the results obtained for the lowest 
NO in harmonic traps where the curvature of the confining potential was 
changed by one order of magnitude. It can be seen that a scaled NO 
$\varphi^0=R^{1/2}\phi^0$ exists, which does not change when any 
parameter of the system is changed keeping 
$\tilde{\rho}$ constant. Even when regions with $n_i=1$ are present 
[Figs.\ \ref{HCB_all}(b),(c)], where the two lowest NO are degenerate, the 
scaled NO exist. Here, the length scale $\zeta$ set by $V_\alpha$ determines 
the length $L=B^{\alpha}_{\tilde{\rho}}\zeta$ over which a nonvanishing 
density is present. $B^{\alpha}_{\tilde{\rho}}$ depends on the 
characteristic density $\tilde{\rho}$ and the power $\alpha$ of the 
confining potential.

\newpage\

\onecolumngrid

\twocolumngrid

The scaling factor $R$ is defined for {\it any} confining potential as 
$R=\sqrt{N_b \zeta/ a}$. A measure of the scaling of the NO  
can be obtained by studying the area below them. 
If the above-mentioned scaling is valid, since this area is expected 
to depend only on $\tilde{\rho}$, we have 
$\int dx | \varphi^0\left( x/\zeta\right) |/\zeta
\sim R^{1/2} (a/\zeta) \sum_i |\phi^0_i |=S_{\tilde{\rho}}$  
implying that $\sum_i |\phi^0_i|=S'_{\tilde{\rho}} \sqrt{N_b}$ 
($S_{\tilde{\rho}}$ and $S'_{\tilde{\rho}}$ depend only on $\tilde{\rho}$). 
Figure \ref{HCB_all}(d) shows the results obtained for $\sum_i |\phi^0_i|$ 
vs $N_b$ (at constant $\tilde{\rho}$) in traps with power $\alpha=8$ 
of the confining potential \cite{foot1}. 
It can be seen that already for $N_b>100$ the 
expected power law $\sqrt{N_b}$ is present, confirming our previous 
statements for large enough systems.

With the findings above, the leading behavior of
the lowest NO $\lambda_0=\sum_{ij}\phi^0_i \rho_{ij}
\phi^0_j$ can be evaluated in the TL and for a given 
$\tilde{\rho}$ as follows. Replacing the sums in $\lambda_0$ by integrals 
($L \gg a$) one obtains
\begin{eqnarray}
\lambda_0 &\sim& (1/a^2)
\int^L_{-L}dx \int^L_{-L}dy 
\frac{\phi^0(x)A^{\alpha}_{\tilde{\rho}}\phi^0(y)}{|(x-y)/a|^{-1/2}} 
\nonumber \\ &=& \left( \zeta/a\right) ^{3/2}R^{-1}
\int^{B^{\alpha}_{\tilde{\rho}}}_{-B^{\alpha}_{\tilde{\rho}}}dX
\int^{B^{\alpha}_{\tilde{\rho}}}_{-B^{\alpha}_{\tilde{\rho}}}dY 
\frac{\varphi^0(X)A^{\alpha}_{\tilde{\rho}}\varphi^0(Y)}{|X-Y|^{-1/2}}
\nonumber \\ &=&C^{\alpha}_{\tilde{\rho}} \sqrt{\zeta/a}
=D^{\alpha}_{\tilde{\rho}}\sqrt{N_b}, \label{lambda0}
\end{eqnarray}
where we did the change of variables $x=X \zeta$, $y=Y \zeta$, and 
$\phi^0=R^{-1/2}\varphi^0$. The integral over $X,Y$ 
depends only on the characteristic density.
Then for a given confining potential with power $\alpha$, 
$C^{\alpha}_{\tilde{\rho}}$ and $D^{\alpha}_{\tilde{\rho}}$ depend only 
on $\tilde{\rho}$, demonstrating that $\lambda_0$ scales in the 
TL as $\sqrt{N_b}$. The same analysis can be done 
with the MDF, where instead of normalizing it by the system size (as usual 
for homogeneous systems) we normalize it by the length scale set by the 
potential [$n_k=(a/\zeta)\sum_{ij} e^{-ik(i-j)}\rho_{ij}$].
Considering the large-$x$ form of the OPDM, and repeating the reasoning above, 
one obtains that $n_{k=0}$ also scales as $\sqrt{\zeta}$ or $\sqrt{N_b}$,
for constant $\tilde{\rho}$, in the TL.

\begin{figure}[h]
\includegraphics[width=0.73\textwidth,height=0.28\textwidth]
{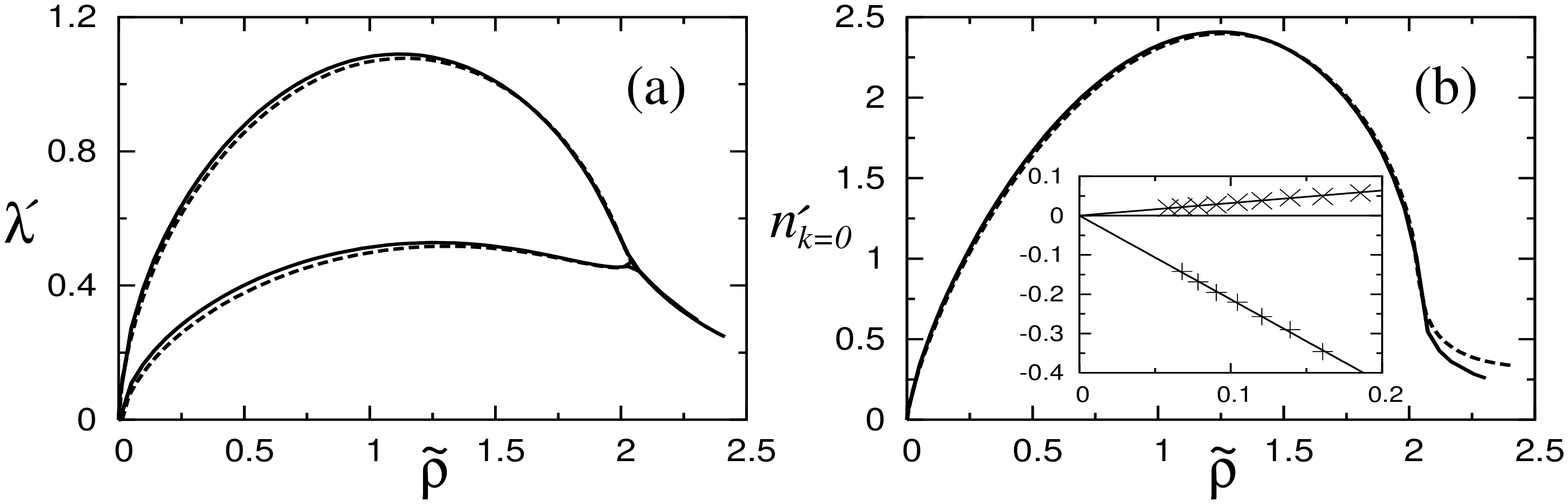}
\caption{$\lambda'$ of the two lowest NO (a) 
and $n'_{k=0}$ (b) vs the characteristic density ($\tilde{\rho}$) 
for traps with power of the confining potential $\alpha=8$,
$V_8a^8=2.0 \times 10^{-19}t$ (continuous line) and 
$V_8a^8=1.0 \times 10^{-15}t$ (dashed line).   
The inset in (b) shows $\delta \lambda'_0$ for $\tilde{\rho}=1.0$ 
($\times$) and $\delta n'_{k=0}$ for $\tilde{\rho}=2.25$ (+) 
vs $\left(\zeta/a\right)^{-1/2}$ (see text); 
the continuous lines show linear behavior.}
\label{Lowest}
\end{figure}

So far we have analyzed the scaling of $\lambda_0$ and $n_{k=0}/\sqrt{\zeta/a}$ 
in the TL. We discuss in what follows its relevance for finite 
size systems. In Fig.\ \ref{Lowest} we plot $\lambda'=\lambda/\sqrt{\zeta/a}$ 
for the first two NO (a) and $n'_{k=0}=n_{k=0}/\sqrt{\zeta/a}$ (b) as a 
function of the characteristic density for two traps with very different
confining potentials with power $\alpha=8$ \cite{foot1}. 
As Fig.\ \ref{Lowest} shows, 
finite size corrections to the leading behavior are very small
since the values of $\lambda$ and $n_{k=0}$ almost scale like in the TL.
Actually, the results for the NO are indistinguishable after the region with 
$n_i=1$ appears in the system, which is the point where the degeneracy sets 
in Fig.\ \ref{Lowest}(a). We find the finite size corrections to $\lambda'$ 
and $n'_{k=0}$ to be 
$\sim 1/\sqrt{\zeta/a}$ so that $\lambda_0/\sqrt{\zeta/a}=
C_{\tilde{\rho}}-E_{\tilde{\rho}}/\sqrt{\zeta/a}$
and $n_{k=0}/\sqrt{\zeta/a}
=F_{\tilde{\rho}}-G_{\tilde{\rho}}/\sqrt{\zeta/a}$,
where $C_{\tilde{\rho}}$, $E_{\tilde{\rho}}$, $F_{\tilde{\rho}}$, 
and $G_{\tilde{\rho}}$ depend only on $\tilde{\rho}$.
In addition  $E_{\tilde{\rho}}$ and $G_{\tilde{\rho}}$ 
can be positive or negative depending on the value of $\tilde{\rho}$. 
As an example we plot in the inset of Fig.\ \ref{Lowest}(b) 
$\delta \lambda'_0=C_{\tilde{\rho}}-
\lambda_0/\sqrt{\zeta/a}$ for $\tilde{\rho}=1.0$ ($E_{\tilde{\rho}}$
is positive) and $\delta n'_{k=0}=F_{\tilde{\rho}}-
n_{k=0}/\sqrt{\zeta/a}$ for $\tilde{\rho}=2.25$ ($G_{\tilde{\rho}}$
is negative). Similar results were obtained for harmonic traps and the 
homogeneous system, in the latter case changing $\zeta/a$ by $N$.

\newpage\

\onecolumngrid

\twocolumngrid

\begin{figure}[h]
\includegraphics[width=0.46\textwidth,height=0.29\textwidth]
{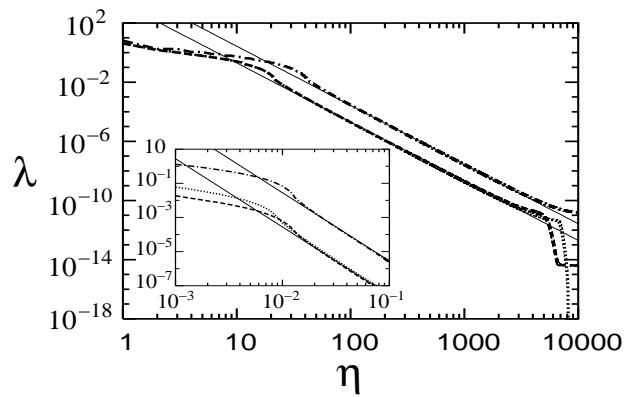}
\caption{Occupation of the NO vs $\eta$ in 
systems with $N=10000$: $N_b=21$ periodic case 
(dashed-dotted line), 
$N_b=11$, $V_2a^2=5.0\times 10^{-12}t$ (dashed line), and  
$N_b=11$, $V_8a^8=5.0\times 10^{-34}t$ (dotted line). The 
inset shows $n_k$ vs $ka$ for the same parameters 
and the same notation. Thin continuous lines correspond to power laws 
$\eta^{-4}$, and $k^{-4}$ in the inset.}
\label{Largealpha}
\end{figure}

Finally, we study the large-$\eta$ behavior of the NO occupations 
($\lambda_\eta$). In contrast to the large-$x$ behavior of the 
density matrix, we do not find a 
universal feature in the large-$\eta$ behavior of $\lambda_\eta$ 
for arbitrary values of the characteristic density $\tilde{\rho}$. 
However, for very small values of $\tilde{\rho}$ we find that a 
universal power law develops in the large-$\eta$ region
of $\lambda_\eta$, as is shown in Fig.\ \ref{Largealpha}. 
The power-law decay is in this case of the form $\lambda_\eta=A_{N_b}/
\eta^4$, where $A_{N_b}$ depends only on the number of particles in 
the trap independently of the confining potential, as is 
shown in Fig.\ \ref{Largealpha}. Since this occurs only for very 
low values of $\tilde{\rho}$ we expect this behavior to be generic 
for the continuous limit. This, to our knowledge, has not been 
discussed before. In the latter limit the high momentum tail of the 
MDF was found to decay as $n_k\sim |k|^{-4}$ for HCB in a harmonic 
trap \cite{minguzzi02}, and for the Lieb-Liniger gas of free and 
harmonically trapped bosons for all values of the interaction strength 
\cite{olshanii03}. Our results for the MDF (inset in Fig.\ 
\ref{Largealpha}) show that the large-$k$ behavior of $n_k$ for 
low $\tilde{\rho}$ is also universal, irrespective of the confining potential.

\begin{figure}[h]
\includegraphics[width=0.45\textwidth,height=0.28\textwidth]
{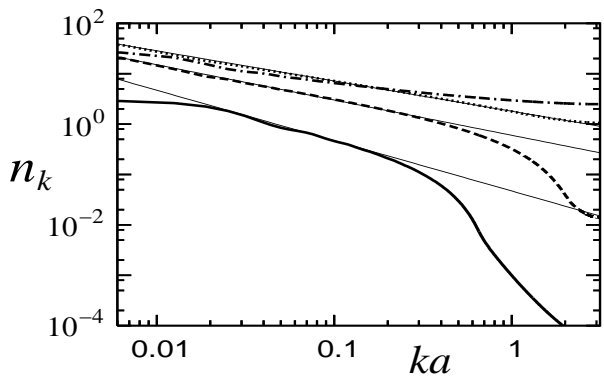}
\caption{MDF for systems with $V_2a^2=3.0\times 10^{-5}t$ and 
$N_b=11$ (continuous line), $N_b=101$ (dashed line), 
and $N_b=401$ (dotted line), $N_b=591$ (dashed-dotted line). 
Accompanying thin continuous lines correspond to power laws: 
$k^{-1}$ for $N_b=11$, $k^{-0.7}$ for $N_b=101$, 
and $k^{-0.6}$ for $N_b=401$.}
\label{nk}
\end{figure}

At this point it is important to remark that the universal behavior 
and scaling relations shown in Figs.\ \ref{Largex}--\ref{Largealpha} 
appear already at moderate number of particles, and hence, are relevant 
for experiments. However, the $|k|^{1/2}$ singularity in $n_k$, 
well known from the homogeneous system, is difficult to see explicitly 
in such situations, in contrast to previous claims \cite{giorgini03}. 
Fitting power laws $n_k\sim k^{-\beta}$ for finite systems could lead to 
wrong conclusions about the large-$x$ behavior of the OPDM, 
as shown in Fig.\ \ref{nk}. For very low fillings ($N_b$=11), 
a ``power-law'' behavior with $\beta$=1 may be seen 
before the $\beta$=4 is established for large $k$. Increasing 
the number of particles leads to a decrease of $\beta$, up to 
$\beta$=0.6 ($N_b$=401). Hence, the power $\beta$ depends strongly 
on the number of particles and cannot be understood as reflecting 
any universal property of the system. Power-law behavior disappears only 
when $n_i$ reaches 1 in parts of the system ($N_b$=591 in Fig.\ \ref{nk}).

In summary, we have shown that quasi-long-range order is present in 
1D HCB on the presence of a lattice, with a universal 
power-law decay of the OPDM, independent of the power of the 
confining potential. Furthermore, we have shown how the occupation of the 
lowest NO and the value of the MDF at zero momentum are determined by the 
large distance behavior of the OPDM. Even in the cases where a region 
with $n_i$=1 builds up in the middle of the system we find that both 
quantities scale proportionally to $\sqrt{N_b}$ (at constant $\tilde{\rho}$). 
A further universal power-law decay has been found for the eigenvalues of 
the OPDM ($\lambda_\eta$) for large values of $\eta$ at low densities 
($\lambda_\eta\sim \eta^{-4}$). It translates into a corresponding power-law 
decay of the MDF ($n_k\sim |k|^{-4}$) at large momenta also independently 
of the power of the confining potential, pointing to scale invariance in the 
ultraviolet limit of the continuous case. 

{\it Note added.} A HCB gas has been realized very recently on 1D 
lattices by Paredes {\it et~al.} \cite{paredes04}.

We are grateful to HLR-Stuttgart (Project DynMet) for allocation of computer
time, and SFB 382 for financial support. We are indebted to F. G\"ohmann
for bringing to our attention Ref.\ \cite{kitanine02}.


\begin{thebibliography}{99}
\bibitem{foot1} This value of $\alpha$ was chosen just to show that 
our analysis is valid for any $\alpha$ and not only 
for a harmonic potential.
\end{thebibliography}
\end{document}